\documentclass[aps,prd,twocolumn,nofootinbib,superscriptaddress]{revtex4-1}

\usepackage{amsmath,amssymb,amsfonts,bm}
\usepackage{graphicx}
\usepackage{hyperref}
\hypersetup{colorlinks=true,allcolors=blue}
\usepackage{braket}

\newcommand{\Tr}{\operatorname{Tr}}

\newcommand{\Ree}{\operatorname{Re}}
\newcommand{\Imm}{\operatorname{Im}}
\newcommand{\Cov}{\operatorname{Cov}}
\newcommand{\Var}{\operatorname{Var}}
\newcommand{\cH}{\mathcal{H}}
\newcommand{\cA}{\mathcal{A}}
\newcommand{\cO}{\mathcal{O}}
\newcommand{\dd}{\mathrm{d}}
\newcommand{\one}{\mathbf{1}}

\begin{document}

\title{Real-time pseudo entropy and modular-Hamiltonian correlations}

\author{Tatsuhiro Misumi}
\email{misumi@phys.kindai.ac.jp}
\affiliation{Department of Physics, Kindai University, Higashi-Osaka, Osaka 577-8502, Japan}


\begin{abstract}
Pseudo entropy is a complex-valued generalization of entanglement entropy defined from a reduced transition matrix. We study the pseudo entropy associated with a real-time transition matrix between an initial pure state and its unitary time evolution. For a subsystem $A$, we show that the short-time behavior of real-time pseudo entropy is governed by the correlation between the physical Hamiltonian $H$ and the modular Hamiltonian $K_A=-\log\rho_A$ of the initial reduced state, $ S_A(t,0)=S_A(0)-it \langle K_A(H-\langle H\rangle)\rangle + \mathcal{O}(t^2)$. For Hermitian dynamics, the initial imaginary response is controlled by the symmetrized covariance of $H$ and $K_A$ with an overall minus sign, while the initial real response is governed by their commutator. Thus the imaginary part of real-time pseudo entropy is not merely a branch artifact: it is a time-oriented modular response generated by the correlation between microscopic time evolution and subsystem coarse graining.  We clarify the relation of this result to the known first law of pseudo entropy, derive an all-order expression in a Schmidt-diagonal model, recover thermal pseudo entropy as a special case, illustrate the covariance/commutator decomposition in a two-qubit model, and confirm the covariance response in transverse-field Ising-chain quenches, including a finite-size study of a modular susceptibility near the Ising critical region. We discuss how this amplitude-level oriented response can be related to ordinary entropy production, and also give a concrete $\mathcal{PT}$-symmetric toy-model illustration of the non-Hermitian extension.
\end{abstract}

\maketitle

\section{Introduction}

The thermodynamic arrow of time is usually formulated as an irreversible growth of entropy after a suitable coarse graining. In a closed Hermitian quantum system, however, the total von Neumann entropy is conserved by unitary time evolution. Irreversibility therefore requires additional structure: a choice of subsystem, environmental decoherence, measurement, coarse graining, or special initial conditions. In stochastic and quantum thermodynamics, this directionality can be quantified by the distinguishability between a forward process and its time reverse, often expressed as a relative entropy \cite{Kawai:2007zz,Parrondo:2009}. This viewpoint has been tested in isolated quantum quenches and extended to quantum systems coupled to general environments \cite{Batalhao:2015,Manzano:2017}.  Related quantum-channel fluctuation relations also show that complex-valued entropy production can encode genuinely quantum symmetry information before a real second-law inequality is recovered \cite{Kwon:2018}. 

These standard formulations are phrased at the level of probabilities, channels, or measurement statistics.  Post-quench entanglement dynamics provides a complementary perspective: even when the global state remains pure under unitary evolution, subsystem entanglement can encode the emergence of local thermodynamics after a quantum quench, especially in integrable systems where quasiparticle propagation gives a quantitative description of entanglement growth \cite{Alba:2016}. It is therefore natural to ask whether there is a more microscopic, amplitude-level entropic quantity that already carries an orientation of time before an ordinary entropy-production inequality is obtained.

Pseudo entropy is particularly suited to this question. It was introduced as a post-selected generalization of entanglement entropy and is defined from a reduced transition matrix rather than from an ordinary reduced density matrix \cite{Nakata:2020luh}. Following the standard construction of Refs.~\cite{Nakata:2020luh,Mollabashi:2020yie,Mollabashi:2021xsd,Goto:2021kln}, given two nonorthogonal states $\ket{\psi}$ and $\ket{\phi}$, one defines
\begin{align}
 \tau^{\psi|\phi}= \frac{\ket{\psi}\bra{\phi}}{\braket{\phi|\psi}},
 \qquad
 \tau_A^{\psi|\phi}=\Tr_{\bar A}\tau^{\psi|\phi},
\end{align}
and the pseudo entropy is defined as
\begin{align}
 S_A^{\psi|\phi}=-\Tr_A\tau_A^{\psi|\phi}\log\tau_A^{\psi|\phi}.
 \label{eq:pseudo_entropy_def}
\end{align}
Since $\tau_A^{\psi|\phi}$ is generically non-Hermitian, $S_A^{\psi|\phi}$ is generically complex. Its basic properties have been studied in holography, free field theories, spin chains, and related field-theoretic settings \cite{Nakata:2020luh,Mollabashi:2020yie,Mollabashi:2021xsd,Goto:2021kln,Nishioka:2021cxe,Mukherjee:2022jac}. The algebraic structure of complex spectra, reality conditions, and amplification phenomena has also been investigated \cite{Ishiyama:2022odv,Guo:2022sfl}. The first-law-like variation of pseudo entropy is known \cite{Mollabashi:2020yie,Mollabashi:2021xsd}: for nearby transition matrices, the first variation is governed by the modular Hamiltonian of the reference state, in direct analogy with the entanglement first law \cite{Bhattacharya:2012mi,Blanco:2013joa}.

The complex nature of pseudo entropy has recently acquired several further interpretations. In de Sitter holography and timelike entanglement entropy, the imaginary part has been argued to encode an emergent time direction \cite{Doi:2022iyj,Doi:2023zaf}. Related and earlier de Sitter extremal-surface and ghost-CFT analyses had already exhibited complex or nonunitary entropy-like structures \cite{Narayan:2015dS,Narayan:2016ghost, Jatkar:2017ghost,Narayan:2017dSent}. This line of work was developed into time-entanglement and reduced time-evolution-operator constructions \cite{Narayan:2022afv,Narayan:2023timepseudo}, including the observation that appending a projection onto an initial state naturally leads to pseudo entropy between that initial state and its time-evolved final state \cite{Narayan:2023timepseudo}. Further dS/CFT and time-entanglement developments include replica constructions of the de Sitter wavefunction, pseudo-Renyi entropies, transition-matrix operators, weak values, and complex time contours \cite{Narayan:2023dS,Nanda:2025ds,Narayan:2026dS}. 

Thermal pseudo entropy is obtained by analytically continuing the inverse temperature of the thermal entropy to a complex value; equivalently, it is the pseudo entropy of transition matrices between thermofield-double states at different times. It is related to the spectral form factor, and its real and imaginary parts obey Kramers--Kronig relations \cite{Caputa:2024lpa}. Real-time aspects of pseudo entropy have also been studied in two-dimensional CFTs for locally excited states \cite{Guo:2022realpseudo}. Further related aspects of timelike entanglement entropy, including the structure and interpretation of its imaginary part, were discussed in Ref.~\cite{Xu:2024yvf}.

Non-Hermitian quantum systems provide another natural setting for complex entropic quantities, especially in $\mathcal{PT}$-symmetric, pseudo-Hermitian, and more general non-unitary dynamics. The modern development of this field builds on the reality of spectra in $\mathcal{PT}$-symmetric Hamiltonians \cite{Bender:1998ke}, the subsequent formulation of $\mathcal{PT}$-symmetric non-Hermitian quantum mechanics \cite{Bender:2007njh}, and pseudo-Hermiticity \cite{Mostafazadeh:2001jk}; for broad reviews see Refs.~\cite{ElGanainy:2018,Ashida:2020dkc}. In many-body and topological settings, non-Hermiticity further leads to new symmetry classifications, exceptional-point topology, and biorthogonal entanglement structures \cite{Kawabata:2019,Bergholtz:2021,Chang:2019enm,Shimizu:2025}. Complementary unit-invariant singular-value entropies for reduced transition matrices in biorthogonal quantum mechanics were developed and compared with eigenvalue-based pseudo entropy in Ref.~\cite{Caputa:2025ugm}.

In this paper, we study the short-time response of real-time pseudo entropy. We consider the transition matrix between an initial pure state and its unitary time evolution,
\begin{align}
 \ket{\Psi}\longrightarrow\ket{\Psi(t)}=e^{-iHt}\ket{\Psi} ,
\end{align}
and define the corresponding pseudo entropy for a subsystem $A$.  Our central result is
\begin{align}
 S_A(t,0)=S_A(0)-it\,
 \langle K_A(H-\langle H\rangle)\rangle+
 \cO(t^2) ,
 \label{eq:main_intro}
\end{align}
where $K_A=-\log\rho_A$ is the modular Hamiltonian of the initial reduced density matrix.  For Hermitian $H$ and $K_A$, this implies
\begin{align}
 \frac{\dd}{\dd t}\Imm S_A(t,0)\bigg|_{t=0}
 &=-\frac{1}{2}
 \left\langle \{K_A-\langle K_A\rangle,H-\langle H\rangle\}\right\rangle ,
 \label{eq:intro_imag}\\
 \frac{\dd}{\dd t}\Ree S_A(t,0)\bigg|_{t=0}
 &=\frac{1}{2i}\left\langle [K_A,H]\right\rangle .
 \label{eq:intro_real}
\end{align}
The imaginary part is therefore generated by the symmetrized covariance between the physical Hamiltonian and the modular Hamiltonian, while the real part is generated by their commutator.

Eq.~\eqref{eq:main_intro} follows by applying the known first-law-like variation of pseudo entropy \cite{Mollabashi:2020yie, Mollabashi:2021xsd} to the reduced transition matrix generated by real-time evolution. Our contribution is to isolate its real-time content and to decompose it into two directly interpretable responses: a forward/backward-odd imaginary response controlled by the symmetrized covariance between the physical Hamiltonian and the modular Hamiltonian, and a real response controlled by their commutator expectation value. The former is not merely an imaginary contribution produced by a logarithmic branch cut of a non-Hermitian eigenvalue; it is already present at infinitesimal time whenever the physical Hamiltonian is correlated with the modular Hamiltonian of the initial reduced state. This distinction gives a concrete microscopic characterization of an oriented entropic response.

The main results are summarized as follows. First, real-time pseudo entropy admits the universal short-time expansion in Eq.~\eqref{eq:main_intro}, valid for an arbitrary Hermitian Hamiltonian when the logarithm is defined on the support of the initial reduced density matrix. Second, this expansion separates into the covariance and commutator formulae in Eqs.~\eqref{eq:intro_imag} and \eqref{eq:intro_real}; hence the imaginary part is a time-oriented modular covariance, whereas the real part is a commutator response. Third, in a Schmidt-diagonal model this response can be resummed to all orders in terms of a complex tilted distribution, and thermal pseudo entropy is recovered in the thermofield-double special case. Fourth, in finite-dimensional examples, including a noncommuting two-qubit model and transverse-field Ising-chain quenches, we confirm that the imaginary response of pseudo entropy is quantitatively governed by the modular covariance; in the Ising chain the covariance response leads to a modular susceptibility that peaks near the critical region. Finally, we formulate the corresponding one-sided biorthogonal transition matrix for non-Hermitian quantum mechanics, show that the same modular-response structure survives with ordinary expectation values replaced by biorthogonal ones, and illustrate how $\mathcal{PT}$ symmetry breaking changes the analytic structure of the corresponding pseudo entropy from a hyperbolic real-spectrum form to a trigonometric, branch-sensitive form.

The paper is organized as follows. In Sec.~\ref{sec:setup}, we define real-time pseudo entropy. In Sec.~\ref{sec:firstlaw}, we derive the modular response formula and its covariance/commutator decomposition. In Sec.~\ref{sec:schmidt}, we solve a Schmidt-diagonal model to all orders and recover thermal pseudo entropy as a special case. In Sec.~\ref{sec:twoqubit}, we present a two-qubit model. In Sec.~\ref{sec:ising}, we give Ising-chain examples. In Sec.~\ref{sec:nh_extension}, we formulate the non-Hermitian one-sided biorthogonal extension and illustrate the analytic change across $\mathcal{PT}$ symmetry breaking. Section~\ref{sec:discussion} discusses the interpretation, limitations, and relation to entropy production. General second-order response formulae and a short derivation of the trace-logarithm variation are given in the Appendices.

\section{Real-time transition matrix}
\label{sec:setup}

Following the standard pseudo-entropy construction \cite{Nakata:2020luh,Mollabashi:2020yie,Mollabashi:2021xsd,Goto:2021kln}, we now specialize the transition matrix to real-time evolution.  Let the total Hilbert space be bipartitioned as
\begin{align}
 \cH=\cH_A\otimes\cH_{\bar A} .
\end{align}
We start from a normalized pure state $\ket{\Psi}$ and define its real-time evolution by
\begin{align}
 \ket{\Psi(t)}=e^{-iHt}\ket{\Psi} .
\end{align}
The forward transition matrix \cite{Doi:2022iyj,Doi:2023zaf,Caputa:2024lpa,Guo:2022realpseudo,Narayan:2023timepseudo} is
\begin{align}
 \tau(t,0)
 = \frac{\ket{\Psi(t)}\bra{\Psi}}{\braket{\Psi|\Psi(t)}} .
 \label{eq:tau_forward}
\end{align}
It obeys $\Tr\tau(t,0)=1$ but is non-Hermitian for generic $t$. The reduced transition matrix is
\begin{align}
 \tau_A(t,0)=\Tr_{\bar A}\tau(t,0) ,
\end{align}
and the real-time pseudo entropy is
\begin{align}
 S_A(t,0)=-\Tr_A\tau_A(t,0)\log\tau_A(t,0) .
 \label{eq:real_time_PE}
\end{align}
At $t=0$, this reduces to the ordinary entanglement entropy of the initial state,
\begin{align}
 S_A(0)=-\Tr_A\rho_A\log\rho_A ,
 \qquad
 \rho_A=\Tr_{\bar A}\ket{\Psi}\bra{\Psi} .
\end{align}
Throughout the paper we assume that $\rho_A$ is full rank on its support, or equivalently that all formulae are evaluated on the support of $\rho_A$. The logarithm is defined by analytic continuation from $t=0$ until an eigenvalue crosses a branch cut.

The backward transition matrix is
\begin{align}
 \tau(0,t)=\frac{\ket{\Psi}\bra{\Psi(t)}}{\braket{\Psi(t)|\Psi}}
 =\tau(t,0)^\dagger .
 \label{eq:tau_backward}
\end{align}
Thus, as long as the logarithmic branch is followed consistently,
\begin{align}
 S_A(0,t)=S_A(t,0)^*  .
 \label{eq:backward_conjugate}
\end{align}
The orientation of the transition is therefore measured by
\begin{align}
 \cA_A(t)=\frac{1}{2i}\left[S_A(t,0)-S_A(0,t)\right] .
 \label{eq:arrow_def}
\end{align}
When Eq.~\eqref{eq:backward_conjugate} holds, $\cA_A(t)=\Imm S_A(t,0)$. This quantity changes sign under the exchange of the forward and backward transition matrices.  We call it the oriented response of the subsystem. The first-law analysis below identifies its leading short-time behavior as a modular-Hamiltonian response.

It is important that this notion of orientation is not identical to the naive replacement $t\to -t$ in Eq.~\eqref{eq:tau_forward}. The operation relevant to pseudo entropy is the exchange of the bra and ket states in the transition matrix.  This is the natural analogue of exchanging a forward process and its reverse.

\section{First-law response of real-time pseudo entropy}
\label{sec:firstlaw}

We now derive the leading short-time behavior of Eq.~\eqref{eq:real_time_PE}. The modular Hamiltonian of the initial reduced state is
\begin{align}
 K_A=-\log\rho_A .
 \label{eq:modular_def}
\end{align}
Throughout this section the logarithm is understood on the support of $\rho_A$.  Equivalently, one may first restrict all reduced operators to this support, or use the regulator $\rho_A\to\rho_A+\epsilon\one_A$ and take $\epsilon\to0$ after the variation when the limit is finite. When inserted in expectation values over the total Hilbert space, $K_A$ means $K_A\otimes\one_{\bar A}$.

Expanding the numerator and denominator of Eq.~\eqref{eq:tau_forward} gives
\begin{align}
 \tau(t,0)
 =\ket{\Psi}\bra{\Psi}
 -it\,(H-\langle H\rangle)\ket{\Psi}\bra{\Psi}
 +\cO(t^2) ,
 \label{eq:tau_expansion}
\end{align}
where $\langle\cdots\rangle=\bra{\Psi}\cdots\ket{\Psi}$. Hence
\begin{align}
 \delta\tau_A
 =-it\,\Tr_{\bar A}\left[(H-\langle H\rangle)
 \ket{\Psi}\bra{\Psi}\right]+\cO(t^2) .
\end{align}
For a trace-preserving variation, the first variation of $-\Tr\rho\log\rho$ is
\begin{align}
 \delta S_A=\Tr_A(\delta\tau_A K_A) .
 \label{eq:variation_entropy}
\end{align}
This gives
\begin{align}
 S_A(t,0)=S_A(0)-it\,C_A+\cO(t^2) ,
 \label{eq:main_first_law}
\end{align}
with
\begin{align}
 C_A=\left\langle K_A(H-\langle H\rangle)\right\rangle .
 \label{eq:CA_def}
\end{align}
Eq.~\eqref{eq:main_first_law} is the real-time specialization of the first-law-like relation for pseudo entropy. What is specific to real time is the factor $-i$ and the consequent separation into an imaginary covariance response and a real commutator response.

For Hermitian $H$ and $K_A$, $C_A$ is generally complex because the two operators need not commute.  Separating Eq.~\eqref{eq:main_first_law} into real and imaginary parts gives
\begin{align}
 \frac{\dd}{\dd t}\Imm S_A(t,0)\bigg|_{t=0}
 &=-\Ree C_A ,
 \label{eq:dIm_C}\\
 \frac{\dd}{\dd t}\Ree S_A(t,0)\bigg|_{t=0}
 &=\Imm C_A .
 \label{eq:dRe_C}
\end{align}
Using
\begin{align}
 \Ree C_A
 &= \frac{1}{2}\left\langle
 \{K_A-\langle K_A\rangle,H-\langle H\rangle\}
 \right\rangle ,
 \label{eq:sym_cov}\\
 \Imm C_A
 &= \frac{1}{2i}\left\langle [K_A,H]\right\rangle ,
 \label{eq:comm}
\end{align}
we obtain
\begin{align}
 \frac{\dd}{\dd t}\Imm S_A(t,0)\bigg|_{t=0}
 &=-\frac{1}{2}\left\langle
 \{K_A-\langle K_A\rangle,H-\langle H\rangle\}
 \right\rangle ,
 \label{eq:main_im_response}\\
 \frac{\dd}{\dd t}\Ree S_A(t,0)\bigg|_{t=0}
 &=\frac{1}{2i}\left\langle [K_A,H]\right\rangle .
 \label{eq:main_re_response}
\end{align}
Eq.~\eqref{eq:main_im_response} is the main result of this work. It states that the imaginary part of real-time pseudo entropy is generated, at leading order, by the symmetrized covariance between the physical Hamiltonian and the modular Hamiltonian. Since $H$ generates microscopic time evolution while $K_A$ encodes the subsystem information structure, this covariance is a precise measure of how the chosen coarse graining responds to real-time evolution. Eq.~\eqref{eq:main_re_response} shows that the real part, in contrast, responds linearly only when the modular and physical Hamiltonians fail to commute in the initial state.

When $[K_A,H]$ has vanishing expectation value, the real part has no linear response, whereas the imaginary part can still have a nonzero linear response. This is the typical situation in time-reversal-symmetric examples with real initial data. The imaginary part is then the leading oriented contribution.

\section{Schmidt-diagonal model and thermal pseudo entropy}
\label{sec:schmidt}

A useful solvable class is obtained when the initial state is written in Schmidt form,
\begin{align}
 \ket{\Phi}=\sum_n \sqrt{p_n}\ket{n}_A\ket{n}_{\bar A} ,
 \quad
 p_n>0 ,
 \quad
 \sum_n p_n=1 ,
 \label{eq:schmidt_state}
\end{align}
and the Hamiltonian is diagonal in this Schmidt sector,
\begin{align}
 H\ket{n}_A\ket{n}_{\bar A}=E_n\ket{n}_A\ket{n}_{\bar A} .
 \label{eq:schmidt_H}
\end{align}
Then
\begin{align}
 \ket{\Phi(t)}=\sum_n \sqrt{p_n}e^{-iE_nt}
 \ket{n}_A\ket{n}_{\bar A} .
\end{align}
The eigenvalues of the reduced transition matrix are
\begin{align}
 \lambda_n(t)=\frac{p_ne^{-iE_nt}}{\chi(t)} ,
 \qquad
 \chi(t)=\sum_m p_m e^{-iE_mt}  .
 \label{eq:lambda_schmidt}
\end{align}
Thus, the pseudo entropy is
\begin{align}
 S_A(t)=-\sum_n\lambda_n(t)\log\lambda_n(t) \, .
\end{align}

It is useful to rewrite this result in terms of a complex tilted distribution.  We set
$ \theta=-it,\,M(\theta)=\sum_n p_n e^{\theta E_n}$ and define
\begin{align}
 \lambda_n(\theta)
 =
 \frac{p_n e^{\theta E_n}}{M(\theta)} .
 \label{eq:tilted_lambda}
\end{align}
For any quantity $O_n$ assigned to the Schmidt label $n$, we define its tilted average by
\begin{align}
 \langle O\rangle_\theta
 =
 \sum_n \lambda_n(\theta) O_n .
 \label{eq:tilted_average}
\end{align}
In particular,
\begin{align}
 \langle E\rangle_\theta
 =
 \sum_n \lambda_n(\theta) E_n .
 \label{eq:tilted_E}
\end{align}
The reduced density matrix at $t=0$ is diagonal in the Schmidt basis,
\begin{align}
 \rho_A
 =
 \sum_n p_n |n\rangle_A {}_A\langle n| .
 \label{eq:rho_schmidt_diagonal}
\end{align}
Therefore the modular Hamiltonian $K_A=-\log\rho_A$ has eigenvalues
\begin{align}
 K_n=-\log p_n .
 \label{eq:schmidt_modular_eigenvalues}
\end{align}
We then define
\begin{align}
 \langle K\rangle_\theta
 =
 \sum_n \lambda_n(\theta) K_n .
 \label{eq:tilted_K}
\end{align}
Thus, in this diagonal model, $E$ and $K$ are diagonal variables taking the values $E_n$ and $K_n$, respectively. The tilted average $\langle\cdots\rangle_\theta$ is a complex-weighted average for $\theta\neq0$, whereas the cumulants appearing below are evaluated at $\theta=0$ with respect to the genuine probabilities $p_n$.

Using
\begin{align}
 \log\lambda_n(\theta)
& =
 \log p_n+\theta E_n-\log M(\theta)
\nonumber\\
&=
 -K_n+\theta E_n-\log M(\theta),
\end{align}
the pseudo entropy becomes
\begin{align}
 S_A(\theta)
=
 \langle K\rangle_\theta
 -
 \theta\langle E\rangle_\theta
 +
 \log M(\theta).
 \label{eq:exact_schmidt}
\end{align}
This formula shows that real-time pseudo entropy is a complex exponential tilt of the entanglement spectrum by the physical energies.

Expanding Eq.~\eqref{eq:exact_schmidt} around $\theta=0$ gives
\begin{align}
 S_A(t)=S_A(0)&-it\,\Cov(K,E)\notag\\
 &+\frac{t^2}{2}\left[\Var(E)-\kappa(K,E,E)\right]
 +\cO(t^3) .
 \label{eq:schmidt_expansion}
\end{align}
Here all averages are taken with respect to the original Schmidt probabilities $p_n$, namely
\begin{align}
 \langle O\rangle=\sum_n p_n O_n  .
\end{align}
In this diagonal model, the variance, covariance, and mixed cumulant are ordinary classical cumulants of the distribution $p_n$,
\begin{align}
 \Var(E)
 &=
 \langle E^2\rangle-\langle E\rangle^2,
 \notag\\
 \Cov(K,E)
 &=
 \langle KE\rangle-\langle K\rangle\langle E\rangle,
 \notag\\
 \kappa(K,E,E)
 &=
 \left\langle
 \bigl(K-\langle K\rangle\bigr)
 \bigl(E-\langle E\rangle\bigr)^2
 \right\rangle .
 \label{eq:var_cov_definitions}
\end{align}
Thus
\begin{align}
 \Imm S_A(t)=-\Cov(K,E)t+\cO(t^3) ,
 \label{eq:schmidt_imag}
\end{align}
while
\begin{align}
 \Ree S_A(t)=S_A(0)+\frac{t^2}{2}\left[\Var(E)-\kappa(K,E,E)\right]
 +\cO(t^4) .
 \label{eq:schmidt_real}
\end{align}
The real part is not constrained to increase; the result is a complex response, not a second-law inequality. We emphasize that the notation $\Cov(K,E)$ in this section refers to the ordinary covariance of two diagonal random variables. In the general noncommuting setting, the corresponding object is the symmetrized modular covariance written explicitly in Eqs.~\eqref{eq:intro_imag} and \eqref{eq:main_im_response}.

Thermal pseudo entropy \cite{Caputa:2024lpa} is recovered by choosing the Schmidt probabilities to be the Boltzmann weights associated with the thermofield-double construction,
\begin{align}
 p_n
 &=
 \frac{e^{-\beta E_n}}{Z(\beta)},
 \qquad
 Z(\beta)=\sum_n e^{-\beta E_n}.
\end{align}
Then
\begin{align}
 K_n=\beta E_n+\log Z(\beta),
\end{align}
so that
\begin{align}
 \Cov(K,E)=\beta\Var(E).
\end{align}
Eq.~\eqref{eq:schmidt_imag} becomes
\begin{align}
 \Imm S_{\rm TPE}(\beta,t)
 =
 -\beta\Var_\beta(H)t+\cO(t^3),
 \label{eq:TPE_shorttime}
\end{align}
where
$\Var_\beta(H)=\langle H^2\rangle_\beta-\langle H\rangle_\beta^2$
is the thermal energy variance. This agrees with the short-time expansion of the thermal pseudo entropy defined in Ref.~\cite{Caputa:2024lpa}, where the reduced transition matrix is the thermal density matrix analytically continued to the complex inverse
temperature $s=\beta+it$.  Indeed, using their definition
$S_{\rm th}(s)=(1-s\partial_s)\log Z(s)$, one finds
$\Imm S_{\rm th}(\beta+it)=-\beta\Var_\beta(H)t+\cO(t^3)$.

\section{Two-qubit example with noncommuting generators}
\label{sec:twoqubit}

We next present a minimal example in which the modular Hamiltonian and the physical Hamiltonian do not commute.  Consider
\begin{align}
 \ket{\Psi}=\sqrt p\ket{00}+e^{i\phi}\sqrt q\ket{11},
 \qquad q=1-p.
 \label{eq:twoqubit_state}
\end{align}
The reduced density matrix of qubit $A$ is
\begin{align}
 \rho_A=p\ket{0}\bra{0}+q\ket{1}\bra{1},
\end{align}
and hence
\begin{align}
 K_A&=K_0\ket{0}\bra{0}+K_1\ket{1}\bra{1},
\end{align}
with $ K_0=-\log p,\,K_1=-\log q$.
We choose
\begin{align}
 H=J\sigma_z^{(1)}\sigma_z^{(2)}
 +g\sigma_x^{(1)}\sigma_x^{(2)}.
 \label{eq:twoqubit_H}
\end{align}
Here $\sigma_\alpha^{(1)}=\sigma_\alpha\otimes \mathbf 1$ and $\sigma_\alpha^{(2)}=\mathbf 1\otimes\sigma_\alpha$ act on the first and second qubits, respectively.
This Hamiltonian preserves the subspace spanned by $\ket{00}$ and $\ket{11}$.  In this subspace it is $H=J\one+g\sigma_x$.

The evolved state is
\begin{align}
 \ket{\Psi(t)}=A(t)\ket{00}+B(t)\ket{11},
\end{align}
with
\begin{align}
 A(t)&=e^{-iJt}\left[\sqrt p\cos(gt)-ie^{i\phi}\sqrt q\sin(gt)\right],\\
 B(t)&=e^{-iJt}\left[e^{i\phi}\sqrt q\cos(gt)-i\sqrt p\sin(gt)\right].
\end{align}
The overlap is
\begin{align}
 \braket{\Psi|\Psi(t)}=e^{-iJt}
 \left[\cos(gt)-i2\sqrt{pq}\cos\phi\sin(gt)\right].
\end{align}
The reduced transition matrix remains diagonal in the $\ket{0},\ket{1}$ basis, with eigenvalues
\begin{align}
 \lambda_0(t)
 &= \frac{p\cos(gt)-i\sqrt{pq}e^{i\phi}\sin(gt)}{
 \cos(gt)-i2\sqrt{pq}\cos\phi\sin(gt)},
 \label{eq:lambda0_twoqubit}\\
 \lambda_1(t)
 &= \frac{q\cos(gt)-i\sqrt{pq}e^{-i\phi}\sin(gt)}{
 \cos(gt)-i2\sqrt{pq}\cos\phi\sin(gt)}.
 \label{eq:lambda1_twoqubit}
\end{align}
Therefore
\begin{align}
 S_A(t)=-\lambda_0(t)\log\lambda_0(t)
 -\lambda_1(t)\log\lambda_1(t).
 \label{eq:twoqubit_entropy}
\end{align}

For this model, $C_A=\langle K_A(H-\langle H\rangle)\rangle$ is
\begin{align}
 C_A=g\sqrt{pq}(K_0-K_1)
 \left[(q-p)\cos\phi+i\sin\phi\right].
 \label{eq:twoqubit_CA}
\end{align}
Thus
\begin{align}
 \frac{\dd}{\dd t}\Imm S_A(t)\bigg|_{t=0}
 &=-g\sqrt{pq}(K_0-K_1)(q-p)\cos\phi,
 \label{eq:twoqubit_imag_slope}\\
 \frac{\dd}{\dd t}\Ree S_A(t)\bigg|_{t=0}
 &=g\sqrt{pq}(K_0-K_1)\sin\phi.
 \label{eq:twoqubit_real_slope}
\end{align}
For a real initial state, $\phi=0$, the real part has no linear response while the imaginary part is generated by the modular covariance. For $\phi\neq0$, the commutator expectation is nonzero and the real part also develops a linear response. This illustrates the distinction between the covariance contribution to $\Imm S_A$ and the commutator contribution to $\Ree S_A$.

\begin{figure}[tbp]
 \centering
 \includegraphics[width=\columnwidth]{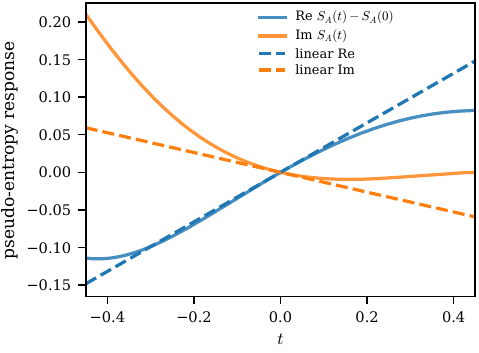}
 \caption{Real and imaginary parts of the real-time pseudo entropy in the two-qubit model. We plot $\Ree S_A(t)-S_A(0)$ and $\Imm S_A(t)$ for $p=0.3$, $q=0.7$, $\phi=\pi/4$, and $g=1.2$. Dashed lines show the linear predictions from Eqs.~\eqref{eq:twoqubit_imag_slope} and \eqref{eq:twoqubit_real_slope}. The imaginary part is governed by the modular covariance, while the real part is governed by the commutator expectation value.}
 \label{fig:twoqubit}
\end{figure}

For example, for $p=0.3$, $q=0.7$, $\phi=\pi/4$, and $g=1.2$, one finds
\begin{align}
 C_A=0.1317868+0.3294671i.
\end{align}
Accordingly,
\begin{align}
 \frac{\dd}{\dd t}\Ree S_A(t)\bigg|_{t=0}&=0.3294671,\notag\\
 \frac{\dd}{\dd t}\Imm S_A(t)\bigg|_{t=0}&=-0.1317868.
\end{align}
These values agree with the direct expansion of Eq.~\eqref{eq:twoqubit_entropy}. These results are shown in Fig.~\ref{fig:twoqubit}.

\section{Many-body quench example}
\label{sec:ising}

\begin{figure}[tbp]
 \centering
 \includegraphics[width=\columnwidth]{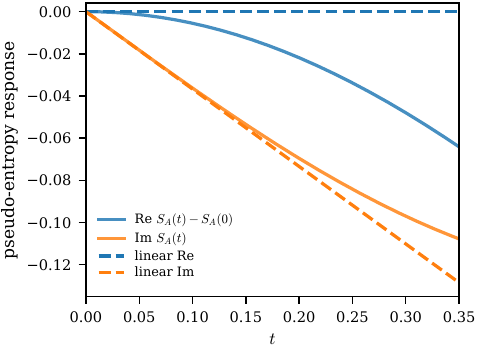}
 \caption{Real-time pseudo entropy after a transverse-field Ising-chain quench.  The system is an open chain with $L=6$, initial field $h_i=1.0$, final field $h_f=1.55$, and subsystem size $L_A=3$. The real part has no linear term for this real time-reversal-symmetric setup, while the imaginary part is linear with slope fixed by the modular covariance. Dashed lines show the first-order predictions.}
 \label{fig:tfim}
\end{figure}

The preceding examples are analytically controlled. To show that the modular covariance response is not restricted to two-level systems or Schmidt-diagonal evolution, we also consider many-body quenches in the open transverse-field Ising chain \cite{Sachdev:2011},
\begin{align}
 H(h)=-\sum_{j=1}^{L-1}\sigma_z^{(j)}\sigma_z^{(j+1)}
 -h\sum_{j=1}^{L}\sigma_x^{(j)}.
 \label{eq:tfim_H}
\end{align}
We first take $L=6$, choose the initial state to be the ground state of $H(h_i)$ with $h_i=1.0$, and evolve it with $H(h_f)$ with $h_f=1.55$. The subsystem $A$ consists of the first three spins.  The pseudo entropy is computed directly from
\begin{align}
 \tau_A(t,0)=\Tr_{\bar A}
 \frac{e^{-iH(h_f)t}\ket{\Psi_0}\bra{\Psi_0}}
 {\bra{\Psi_0}e^{-iH(h_f)t}\ket{\Psi_0}}.
\end{align}
Since the Hamiltonian matrices and the initial ground state can be chosen real in the computational basis, the expectation value of the commutator term vanishes in Eq.~\eqref{eq:main_re_response}. Thus the linear response of the real part is absent, although the modular Hamiltonian and the final Hamiltonian need not commute as operators.
The leading oriented response is therefore entirely imaginary. Numerically, we find
\begin{align}
 &S_A(0)=0.327988,\nonumber\\
 &C_A =\langle K_A(H(h_f)-\langle H(h_f)\rangle)\rangle
 =0.367077,
\end{align}
so that
\begin{align}
 \frac{\dd}{\dd t}\Imm S_A(t,0)\bigg|_{t=0}
 &=-0.367077,\notag\\
 \frac{\dd}{\dd t}\Ree S_A(t,0)\bigg|_{t=0}&=0.
\end{align}
The direct exact-diagonalization result is shown in Fig.~\ref{fig:tfim}. The imaginary part follows the covariance prediction at short times, while the real part starts quadratically.  This example illustrates that the imaginary modular response persists in a genuinely many-body reduced transition matrix with a nontrivial entanglement spectrum.

The same computation also gives a useful finite-size diagnostic. Let the initial state be the ground state of $H(h)$ and consider an infinitesimal quench $h\to h+\delta h$. Since the initial state is an eigenstate of $H(h)$, the response to the unquenched Hamiltonian vanishes and the leading slope is controlled by the perturbation
\begin{align}
 V=\partial_h H(h)=-\sum_{j=1}^{L}\sigma_x^{(j)}.
\end{align}
We define the modular susceptibility
\begin{align}
 \chi_A(h)=\Ree\left\langle K_A\left(V-\langle V\rangle\right)\right\rangle_{h},
 \label{eq:modular_susceptibility}
\end{align}
where the expectation value and $K_A=-\log\rho_A$ are evaluated in the ground state of $H(h)$. For a small quench, Eq.~\eqref{eq:main_im_response} gives
\begin{align}
 \frac{\dd}{\dd t}\Imm S_A(t,0)\bigg|_{t=0}
 =-\delta h\,\chi_A(h)+\cO(\delta h^2).
 \label{eq:susceptibility_slope}
\end{align}
Thus $\chi_A$ measures how sensitively the oriented pseudo-entropy response reacts to the transverse-field perturbation. It is closely analogous in spirit to fidelity susceptibility and the quantum-geometric approach to quantum phase transitions, where the response of the ground-state overlap to a parameter deformation detects criticality \cite{Zanardi:2006,Gu:2010}. The difference is that $\chi_A(h)$ is not a global overlap susceptibility: it is a subsystem and modular susceptibility, correlating the perturbing operator $V=\partial_h H$ with the entanglement Hamiltonian $K_A$. It therefore probes how the local information structure of the chosen subsystem responds to a Hamiltonian deformation.

\begin{figure}[tbp]
 \centering
 \includegraphics[width=\columnwidth]{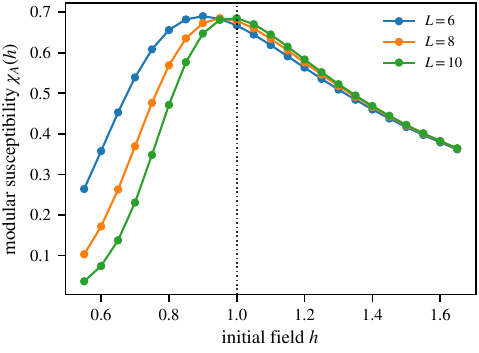}
 \caption{Finite-size modular susceptibility $\chi_A(h)$ in the open transverse-field Ising chain, Eq.~\eqref{eq:modular_susceptibility}, for $L=6,8,10$ and $L_A=L/2$.  The vertical dotted line marks the thermodynamic critical point $h=1$.  The peak sharpens and moves toward the critical region as $L$ increases, suggesting that the imaginary pseudo-entropy slope probes many-body critical correlations through the modular covariance.}
 \label{fig:tfim_size}
\end{figure}

Fig.~\ref{fig:tfim_size} shows exact-diagonalization data for open chains with $L=6,8,10$ and $L_A=L/2$.  The susceptibility develops a broad maximum around the finite-size critical region and the curves become sharper as the system size is increased.  We do not attempt a scaling collapse here, since the accessible sizes are small and open-boundary effects are visible.  Nevertheless, the data indicate that the modular covariance is a many-body response function with nontrivial finite-size dependence.

\section{Non-Hermitian extension and \texorpdfstring{$\mathcal{PT}$}{PT}-symmetric toy model}
\label{sec:nh_extension}

We finally comment on how the present construction extends to non-Hermitian quantum mechanics. The essential point is that pseudo entropy is a transition quantity. It is defined not from an ordinary density matrix of a single state, but from a transition matrix between two boundary states. This distinction becomes especially important in non-Hermitian systems, where right and left states are independent objects. For a non-Hermitian Hamiltonian $H\neq H^\dagger$, right and left eigenstates are defined by
\begin{align}
 H\ket{R_n}=E_n\ket{R_n},
 \qquad
 \bra{L_n}H=E_n\bra{L_n}.
 \label{eq:nh_left_right_eigenstates}
\end{align}
Thus a natural transition amplitude involves both a right state and a left state. The question relevant for pseudo entropy is therefore not only how the entropy of a non-Hermitian state evolves, but how to define a transition matrix appropriate for a pair of boundary states.

Let $\ket R$ and $\bra L$ be a pair of right and left states normalized as
\begin{align}
 \braket{L|R}=1.
\end{align}
The direct non-Hermitian analogue of the real-time transition matrix used in the preceding sections is
\begin{align}
 \ket{R(t)}=e^{-iHt}\ket R,
 \qquad
 \tau^{RL}(t,0)=
 \frac{\ket{R(t)}\bra L}{\braket{L|R(t)}} .
 \label{eq:nh_transition}
\end{align}
Here the right boundary state is evolved, while the left boundary state is kept fixed. This asymmetric construction is the transition-matrix analogue of the pseudo-entropy construction: it represents a pre- and post-selected amplitude, or equivalently a transition amplitude with asymmetric boundary conditions. A complementary singular-value-based approach to such biorthogonal transition matrices was developed in Ref.~\cite{Caputa:2025ugm}, where unit-invariant SVD entropies were constructed and compared with eigenvalue-based pseudo entropy.

The reduced transition matrix and the corresponding modular Hamiltonian are defined by
\begin{align}
 \tau_A^{RL}(t,0)
 &=
 \Tr_{\bar A}\tau^{RL}(t,0),
 \notag\\
 \rho_A^{RL}
 &=
 \tau_A^{RL}(0,0),
 \notag\\
 K_A^{RL}
 &=
 -\log \rho_A^{RL}.
 \label{eq:nh_reduced_transition}
\end{align}
Expanding Eq.~\eqref{eq:nh_transition} at short times gives
\begin{align}
 \tau^{RL}(t,0)
 &=
 \rho^{RL}
 -it\,
 \bigl(H-\langle H\rangle_{RL}\bigr)\rho^{RL}
 +\cO(t^2),
 \notag\\
 \rho^{RL}
 &=
 \ket R\bra L,
 \label{eq:nh_transition_expansion}
\end{align}
where the biorthogonal expectation value is
\begin{align}
 \langle O\rangle_{RL}
 =
 \bra L O\ket R .
 \label{eq:nh_expectation}
\end{align}
Applying the same trace-logarithm variation as in the Hermitian case, one obtains
\begin{align}
 S_A^{RL}(t,0)
 =
 S_A^{RL}(0)
 -it
 \left\langle
 K_A^{RL}
 \bigl(H-\langle H\rangle_{RL}\bigr)
 \right\rangle_{RL}
 +\cO(t^2).
 \label{eq:nh_response}
\end{align}
Therefore the initial complex response of the non-Hermitian pseudo entropy is governed by a biorthogonal modular correlator.  If we define
\begin{align}
 C_A^{RL}
 =
 \left\langle
 K_A^{RL}
 \bigl(H-\langle H\rangle_{RL}\bigr)
 \right\rangle_{RL},
 \label{eq:nh_CA}
\end{align}
then
\begin{align}
 \frac{\dd}{\dd t}
 \Ree S_A^{RL}(t,0)
 \bigg|_{t=0}
 &=
 \Imm C_A^{RL},
 \notag\\
 \frac{\dd}{\dd t}
 \Imm S_A^{RL}(t,0)
 \bigg|_{t=0}
 &=
 -\Ree C_A^{RL}.
 \label{eq:nh_reim}
\end{align}
For genuinely non-Hermitian dynamics, $C_A^{RL}$ is a biorthogonal modular correlator and need not reduce to the symmetrized covariance of two Hermitian operators. Therefore its real and imaginary parts should not be interpreted in the same way as in the Hermitian covariance/commutator decomposition. In a $\mathcal{PT}$-unbroken or, more generally, pseudo-Hermitian regime, an appropriate metric may render the relevant spectrum and expectation values real \cite{Mostafazadeh:2001jk,Bender:2007njh}, and Eq.~\eqref{eq:nh_response} becomes a close analogue of the Hermitian modular-response formula.  In a $\mathcal{PT}$-broken regime or near exceptional points, the eigenvalues of $\rho_A^{RL}$ and the branches of $\log\rho_A^{RL}$ may become singular or multi-valued. This gives a natural setting where an oriented imaginary response, logarithmic branch contributions, and real amplification or decay can coexist. This viewpoint is consistent with recent studies of symmetry, topology, biorthogonal entanglement, and complex entanglement entropy in non-Hermitian many-body systems and complex conformal field theories \cite{Kawabata:2019,Bergholtz:2021,Chang:2019enm,Shimizu:2025}.

It is important to distinguish the transition matrix in Eq.~\eqref{eq:nh_transition} from the more familiar biorthogonal density matrix in which both the right and left states are time evolved. If
\begin{align}
 \ket{R(t)}=e^{-iHt}\ket R,
 \qquad
 \bra{L(t)}=\bra L e^{iHt},
 \label{eq:nh_two_sided_evolution}
\end{align}
one may define
\begin{align}
 \varrho^{RL}(t)
 =
 \frac{\ket{R(t)}\bra{L(t)}}{\braket{L(t)|R(t)}} .
 \label{eq:two_sided_biorthogonal}
\end{align}
For a time-independent Hamiltonian this object obeys
\begin{align}
 \dot\varrho^{RL}(t)
 =
 -i[H,\varrho^{RL}(t)].
 \label{eq:two_sided_equation}
\end{align}
It describes the usual biorthogonal state dynamics. By contrast, Eq.~\eqref{eq:nh_transition} describes an oriented transition amplitude with one boundary state fixed. The two constructions therefore answer different questions. The two-sided object in Eq.~\eqref{eq:two_sided_biorthogonal} is appropriate for state entropy and biorthogonal entanglement dynamics, whereas the one-sided object in Eq.~\eqref{eq:nh_transition} is appropriate for transition entropy and pseudo-entropy response. In particular, the covariance-type term in Eq.~\eqref{eq:nh_response} should not be directly identified with macroscopic entropy production of a non-Hermitian state without specifying an additional measurement, post-selection, or coarse-graining protocol. Clarifying this relation is an important future problem.

A minimal $\mathcal{PT}$-symmetric spectral toy model illustrates how the analytic structure changes across $\mathcal{PT}$ breaking. Consider
\begin{align}
 H_{\rm PT}
 =
 g\sigma_x+i\gamma\sigma_z,
 \qquad
 E_\pm=\pm\Delta,
 \qquad
 \Delta=\sqrt{g^2-\gamma^2}.
 \label{eq:pt_toy}
\end{align}
In a Schmidt-diagonal or thermal pseudo-entropy reduction, we assign
Boltzmann-type weights proportional to $e^{-\beta E_\pm}$, where
$\beta$ is a real inverse-temperature parameter. The complex parameter
entering the pseudo entropy is then
\begin{align}
 s=\beta+it .
\end{align}
Thus, the pseudo entropy is
\begin{align}
 S_{\rm PT}(s)
 =
 \log\bigl(2\cosh s\Delta\bigr)
 -s\Delta\tanh(s\Delta).
 \label{eq:pt_entropy}
\end{align}
For $|\gamma|<|g|$, $\Delta$ is real and the response is the Hermitian two-level result,
\begin{align}
 \Imm S_{\rm PT}(\beta,t)
 =
 -\beta\Delta^2
 \operatorname{sech}^2(\beta\Delta)\,t
 +\cO(t^3).
 \label{eq:pt_unbroken_response}
\end{align}
For $|\gamma|>|g|$, writing $\Delta=i\kappa$, one finds
\begin{align}
 S_{\rm PT}(s)
 =
 \log\bigl(2\cos\kappa s\bigr)
 +\kappa s\tan(\kappa s).
 \label{eq:pt_broken_entropy}
\end{align}
The zeros of $\cos\kappa s$ then generate logarithmic singularities at
\begin{align}
 \kappa s
 =
 \frac{\pi}{2}+\pi n,
 \qquad
 n\in\mathbb Z.
 \label{eq:pt_zeros}
\end{align}
Thus $\mathcal{PT}$ breaking moves the analytic structure of the complex entropy from hyperbolic thermal behavior to trigonometric branch-sensitive behavior. This elementary example captures the mechanism expected near exceptional points in genuine biorthogonal many-body systems \cite{Bender:1998ke,ElGanainy:2018,Ashida:2020dkc,Bergholtz:2021}.\footnote{A complementary nonperturbative perspective on $\mathcal{PT}$-symmetric spectra and $\mathcal{PT}$ breaking is provided by exact-WKB and resurgence analyses. In particular, exact quantization conditions and Stokes automorphisms in $\mathcal{PT}$-symmetric quantum mechanics were studied in Ref.~\cite{Kamata:2024ewkb}, and related exact-WKB methods were applied to resonance, anti-resonance, and $\mathcal{PT}$-breaking sectors in Ref.~\cite{KamataMisumi:2026ewkb}.}

\section{Summary and Discussion}
\label{sec:discussion}

We have identified the leading imaginary part of real-time pseudo entropy as a modular covariance response. What was known is that pseudo entropy has a first-law-like variation controlled by the modular Hamiltonian. What is established here is that, for real-time transition matrices, this variation splits into a time-oriented imaginary response and a commutator-driven real response. The imaginary response is
\begin{align}
 \frac{\dd}{\dd t}\Imm S_A(t,0)\bigg|_{t=0}
 =-\frac{1}{2}\langle\{\Delta K_A,\Delta H\}\rangle,
\end{align}
with $\Delta O =O-\langle O\rangle$
and therefore directly measures the correlation between microscopic time evolution and the modular structure of the subsystem.

The present result suggests a microscopic route by which irreversibility may emerge from unitary quantum dynamics. Although the von Neumann entropy of a closed global system remains constant, the reduced transition amplitudes of a subsystem already distinguish the forward and backward orientations through the quantity $\cA_A(t)$ defined in Eq.~\eqref{eq:arrow_def}. This forward/backward-odd response arises before a measurement, environmental decoherence, or coarse graining has converted quantum amplitudes into classical or quantum-channel probabilities, and its leading behavior is universally governed by the modular covariance derived above. Conventional entropy production is usually formulated at the subsequent probabilistic level, often as a relative entropy between forward and backward processes \cite{Kawai:2007zz,Parrondo:2009,Batalhao:2015,Manzano:2017}. The oriented imaginary pseudo-entropy response may therefore be viewed as a phase-sensitive, amplitude-level precursor of ordinary entropy production and thermodynamic irreversibility. Establishing an explicit coarse-graining or measurement protocol that maps this modular response to a conventional entropy-production functional is an important direction for future work. This interpretation is consistent with quantum fluctuation-theorem formulations, in which entropy production depends on the specified measurement scheme, environment, or recovery map, and is also reminiscent of complex-valued entropy production in quantum-channel fluctuation relations, where the imaginary part encodes genuinely quantum symmetry information before a real second-law inequality emerges \cite{Kwon:2018}.

The imaginary response identified in this paper should also be distinguished from branch-cut imaginary parts. Because $\tau_A(t,0)$ is non-Hermitian, its eigenvalues can cross a branch cut of the logarithm. In addition, the denominator of Eq.~\eqref{eq:tau_forward},
\begin{align}
 G(t)=\braket{\Psi|\Psi(t)},
\end{align}
is precisely the Loschmidt amplitude. Zeros of this amplitude are the standard source of nonanalyticities in the rate function of Loschmidt echoes and underlie dynamical quantum phase transitions, as first emphasized for the transverse-field Ising model in Ref.~\cite{Heyl:2013}. In the present transition-matrix language, such zeros make the normalization singular and can induce logarithmic branch phenomena in pseudo entropy. These effects are important and may be related to Stokes phenomena, dynamical quantum phase transitions, or exceptional-point physics, but they need not be odd under the forward/backward exchange. By contrast, the covariance term in Eq.~\eqref{eq:main_im_response} is tied directly to the orientation of real-time evolution already at infinitesimal time, before a Loschmidt zero or a branch crossing is reached.

The non-Hermitian extension in Sec.~\ref{sec:nh_extension} shows that the modular-response formula is not tied to Hermiticity itself, but to the one-sided transition-matrix structure. In pseudo-Hermitian or $\mathcal{PT}$-unbroken regimes \cite{Mostafazadeh:2001jk, Bender:2007njh}, the response closely parallels the Hermitian covariance formula after replacing ordinary expectation values by biorthogonal ones. In $\mathcal{PT}$-broken regimes, the same formalism displays additional analytic structure: oriented imaginary response, logarithmic branch contributions, and real amplification or decay can no longer be cleanly separated. This is why the non-Hermitian case is best viewed as a natural extension of the present result, rather than as an additional assumption in its derivation.

Several directions remain open. The second-order formula in Appendix~\ref{app:second_order} suggests a direct connection to the pseudo-entropy analogue of quantum Fisher information or relative entropy. It would be useful to rewrite this expression in terms of modular-flow correlators, especially in quantum field theory. The finite-size Ising data presented here suggest, but do not yet establish, a universal scaling interpretation of the modular susceptibility near criticality. Larger systems would be needed for that purpose. It would also be useful to make the coarse-graining map from the oriented modular response to conventional entropy production explicit in open-system or measurement-based protocols. These questions will be addressed elsewhere.

It would also be interesting to understand the present modular-covariance response holographically. Holographic entanglement entropy relates boundary entanglement to bulk extremal surfaces through the Ryu--Takayanagi prescriptions and their quantum and covariant extensions \cite{Ryu:2006bv,Ryu:2006ef,Hubeny:2007xt,Nishioka:2009un,Rangamani:2016dms,Lewkowycz:2013nqa,Faulkner:2013ana,Wall:2012uf,Engelhardt:2014gca,Dong:2013qoa}. Moreover, the entanglement first law, relative entropy, modular Hamiltonians, and entanglement-wedge reconstruction play central roles in deriving bulk gravitational dynamics from boundary entanglement \cite{Bhattacharya:2012mi,Blanco:2013joa,Lashkari:2013koa,Faulkner:2013ica,Jafferis:2015del,Almheiri:2014lwa,Dong:2016eik}. Since holographic pseudo entropy is already known to be related to complex or Euclidean time-dependent extremal areas \cite{Nakata:2020luh,Doi:2022iyj,Doi:2023zaf,Caputa:2024lpa}, the formula derived here suggests that the leading imaginary part of such complex extremal areas should have a boundary interpretation as a modular covariance with the real-time Hamiltonian.

This perspective is particularly close to the de Sitter extremal-surface and time-entanglement program of Refs.~\cite{Narayan:2015dS,Narayan:2016ghost,Jatkar:2017ghost,Narayan:2017dSent,Narayan:2022afv,Narayan:2023timepseudo,Narayan:2023dS,Nanda:2025ds,Narayan:2026dS}. In these works, complex de Sitter extremal-surface areas, reduced time-evolution operators, projection-induced pseudo entropies, and pseudo-Renyi quantities from the de Sitter wavefunction were studied from complementary viewpoints. It would be useful to understand whether the modular covariance identified in the present work gives a boundary first-law interpretation of the leading real-time response in these dS/CFT and time-entanglement settings. In particular, the recent result of Ref.~\cite{Fujiki:2025rtx}, which argues that the first law of holographic pseudo entropy is equivalent to the perturbative Einstein equation in three-dimensional de Sitter space, suggests a promising direction for relating the present real-time modular response to gravitational dynamics.

More broadly, the present result gives a universal physical interpretation to the leading imaginary part of real-time pseudo entropy. By identifying it with a covariance between the physical Hamiltonian and the subsystem modular Hamiltonian, it places complex entropic responses within a general framework of modular response. Because this forward/backward-odd response is already present at the level of quantum transition amplitudes, before measurement, environmental decoherence, or coarse graining converts them into probabilistic processes, it may provide a microscopic precursor of ordinary entropy production and thermodynamic irreversibility. The same structure connects finite-dimensional quantum systems, many-body quenches, thermal pseudo entropy, non-Hermitian dynamics, and holographic settings. Real-time pseudo entropy may therefore serve as a general diagnostic of how microscopic quantum dynamics is organized by subsystem information, while providing a possible bridge between complex quantum amplitudes and emergent macroscopic arrows of time.

\begin{acknowledgments}
The author would like to thank Kotaro Tamaoka for reading the present manuscript and giving valuable comments.
The author also thanks Genki Tanaka for fruitful discussions on the present work. The present work originated from discussions arising from the undergraduate thesis research of Genki Tanaka in 2023, who was formerly a student of the author.
The author appreciates the workshop ``Complexification 2024'' at Saga University for providing opportunities for useful discussions. 
This work was supported by Japan Society for the Promotion of Science (JSPS) KAKENHI Grant No.~23K03425 and 22H05118.
\end{acknowledgments}

\appendix

\section{General second-order expansion}
\label{app:second_order}

Here we record the general second-order expansion around $t=0$.  Let
\begin{align}
 \tau_A(t,0)&=\rho_A+tX_A+t^2Y_A+\cO(t^3),\label{eq:app_tau_exp}\\
 \Tr_A X_A&=\Tr_A Y_A=0.
\end{align}
For the real-time transition matrix in Eq.~\eqref{eq:tau_forward},
\begin{align}
 X_A&=-i\Tr_{\bar A}\left[\Delta H\,|\Psi\rangle\langle\Psi|\right],\label{eq:X_A}\\
 Y_A&=\Tr_{\bar A}\left[
 \left(-\frac{1}{2}(\Delta H)^2+\frac{1}{2}\langle(\Delta H)^2\rangle\right)
 |\Psi\rangle\langle\Psi|\right]
 \label{eq:Y_A}
\end{align}
with $\Delta H=H-\langle H\rangle$.
Define the integral kernel
\begin{align}
 \Omega_{\rho_A}^{-1}(X)=\int_0^\infty \dd s\,
 (\rho_A+s)^{-1}X(\rho_A+s)^{-1}.
 \label{eq:Omega_def}
\end{align}
This is the standard Fr\'echet derivative of the matrix logarithm, but we introduce it only as a compact notation for the noncommuting second-order term. Then the entropy expansion is
\begin{align}
 S_A(t,0)&=S_A(0)+t\Tr_A(X_AK_A)\notag\\
 &+t^2 \left[\Tr_A(Y_AK_A)
 -\frac{1}{2}\Tr_A\left(X_A\Omega_{\rho_A}^{-1}(X_A)\right)\right]
 \notag\\
 &+\cO(t^3).
 \label{eq:second_order_general}
\end{align}
This formula is the noncommuting analogue of Eq.~\eqref{eq:schmidt_expansion}. The second term in the square brackets is the analytic continuation of the Bogoliubov-Kubo-Mori quadratic form. For Hermitian density-matrix variations it is the object that appears in the second variation of relative entropy and quantum Fisher information. In the present problem $X_A$ is generically non-Hermitian, so the same bilinear form controls the complex second-order response of pseudo entropy.

In the Schmidt-diagonal case, $X_A$ and $Y_A$ commute with $\rho_A$.  Eq.~\eqref{eq:second_order_general} then reduces immediately to Eq.~\eqref{eq:schmidt_expansion}. Thus the cumulant expression in Sec.~\ref{sec:schmidt} is the diagonal limit of the general noncommuting response.

\section{Trace-logarithm variation}
\label{app:variation}

For completeness we recall the variation used in Eq.~\eqref{eq:variation_entropy}.  Let $\rho$ be full rank on its support and let $\delta\rho$ be a trace-preserving infinitesimal variation. The first variation of
\begin{align}
 S(\rho)=-\Tr\rho\log\rho
\end{align}
is
\begin{align}
 \delta S=-\Tr\delta\rho\log\rho-\Tr\rho\,\delta\log\rho.
\end{align}
Using the standard integral representation of the logarithmic variation,
\begin{align}
 \delta\log\rho=\int_0^\infty \dd s\,(\rho+s)^{-1}\,\delta\rho\,(\rho+s)^{-1},
\end{align}
one finds
\begin{align}
 \Tr\rho\,\delta\log\rho=\Tr\delta\rho.
\end{align}
For trace-preserving variations, $\Tr\delta\rho=0$, and hence
\begin{align}
 \delta S=\Tr\delta\rho\,K,
 \qquad
 K=-\log\rho.
\end{align}
The same formula applies to the analytic continuation of the reduced transition matrix near $t=0$, as long as no eigenvalue crosses the chosen logarithmic branch cut.


\end{document}